\documentclass[preprint,showkeys,aps,prb]{revtex4}
\usepackage{amsmath}
\usepackage[dvips]{graphicx}
\usepackage{setspace}

\begin{document}

\title{
Fluctuation Theorem and Central Limit Theorem for the Time-Reversible Nonequilibrium Baker Map
}

\author{
William Graham Hoover           \\
Ruby Valley Research Institute  \\
Highway Contract 60, Box 601    \\
Ruby Valley, NV 8983            \\
}

\date{\today}


\vspace{0.1cm}

\begin{abstract}
The nonequilibrium Time-Reversible Baker Map provides simple illustrations of the
Fluctuation Theorem, the Central Limit Theorem, and the Biased Random Walk.  This is
material in preparation for the Book form of Carol's and my 2016 Kharagpur Lectures.
Comments welcome.
\end{abstract}

\maketitle

\section{Introduction}

In 1993 Denis Evans, Eddie Cohen, and Gary Morriss discovered an interesting symmetry, by now
``well-known'', in their studies of the periodic shear flow of 56 hard disks\cite{b1}.  They kept
track of the time-averaged Gibbs' entropy changes associated with the flow as a function of the
averaging time $\tau$.  At a strainrate $\dot \epsilon$ and over a time window $\tau$ the
distribution of entropy production rates approaches a smooth curve with a mean value,
$\langle \ (\dot S/k)= (V/T)\eta \dot \epsilon^2 \ \rangle_\tau $.  The fluctuations about this
mean necessarily satisfy the Central Limit Theorem for large $\tau$.

Evans, Cohen, and Morriss stressed that both positive and {\it negative} values of the entropy
production can be observed if the system is not too small ( 56 soft disks in their case ) and $\tau$
is not too large ( a few collision times ).  At equilibrium the positive and negative values even
out over time.  Away from equilibrium the positive values win out. The relatively few time intervals
with negative values correspond to periods of entropy decrease.  These unlikely fluctuations are
reflected in the title of their paper, ``Probability of Second Law Violations in Shearing Steady
Flows''.  Their key insight was the definition of a nonequilibrium measure in terms of the local
Lyapunov exponents.

I expected that an analog of their shear-stress fluctuations could be found in the qualitatively
simpler dynamics of a nonequilibrium Baker Map.\cite{b2,b3,b4}  The simplicity of the map provides
a worthwhile pedagogical approach to understanding the Evans-Cohen-Morriss ``Fluctuation Theorem''.
The ``Theorem'' is not quite an identitity.  It relates the probability of phase-space trajectory
fragments forward-in-time for a time interval $\tau$ to the less-likely probability of observing
these same fragments with the time order reversed :
$$
[ \ {\rm prob}_{\rm forward}(\tau)/{\rm prob}_{\rm backward}(\tau) \ ]
\simeq e^{\tau \dot S/k} \simeq e^{\Delta S/k} \ [ \ {\rm Fluctuation \ Theorem} \ ] \ .
$$
I have used $\simeq$ rather than = as reminders that the equalities describe limiting cases and are
not valid for short sampling times, $\tau \simeq 0$.

For long times the theorem corresponds to a comparison of the zero-probability unstable repellor to
the probability-one strange attractor generated by the forward mapping.  It is interesting that
Reference 7 of Evans-Cohen-Morriss' paper describes the measurement of fluctuations for a ( fractal )
``Skinny Baker Map''\cite{b5}. The classic Baker Map maps the unit square $(0 < x,y < 1)$ onto itself
with no change in area ( a caricature of Liouville's phase volume conservation at equilibrium.
We consider a generalization of that map to a $2 \times 2$ square, centered on the origin and rotated
$45^o$ so as to fit the conventional definition of time reversibility for the coordinate $q$ and
momentum $p$ :
$$
q = (\sqrt{(1/2)}(dx + dy) \ ; \ p =  (\sqrt{(1/2)}(dx - dy) \ {\rm for}
$$
$$
dx = 2x - 1 \ ; \ dy = 2y - 1 \ ; \ -1 < (dx,dy) < +1 \ .     
$$
The expansion and rotation, together with defining $(q,p)$ coordinates relative
to the center of the square, gives a diamond-shaped time-reversible ergodic dissipative map $B$ :
$$
B(q,p) = (q_{\rm new},p_{\rm new}) \longleftrightarrow B(q_{\rm new},-p_{\rm new}) = (q,-p) \ .
$$

\begin{figure}[h]
\centerline{\includegraphics[width=4.in,angle=-90,bb= 26 11 581 784]{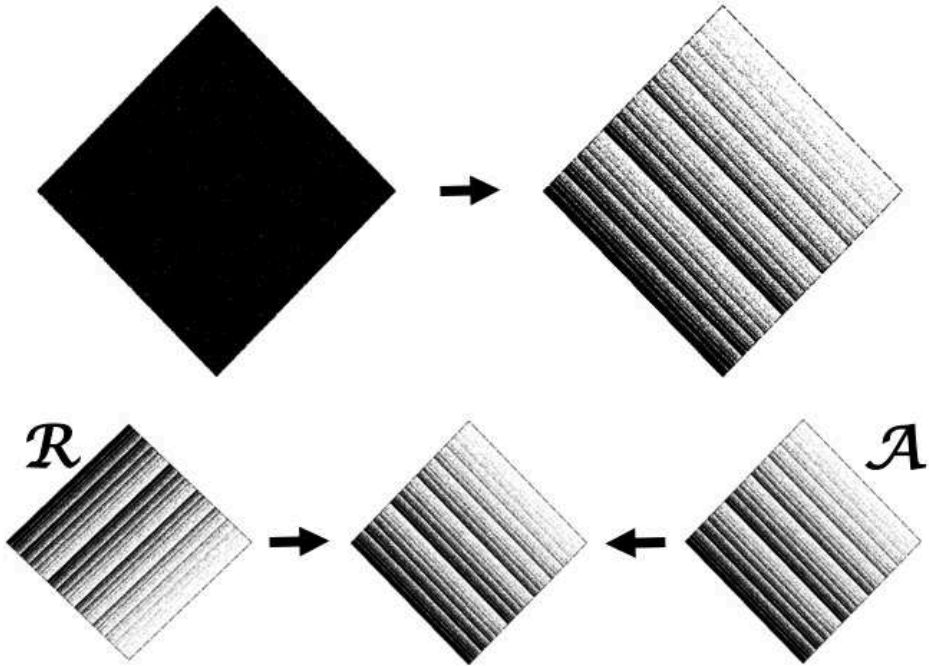}}
\caption
{
Here I show ( at top left ) 250,000 points generated for the equilibrium Baker Map in which
areas are unchanged by the mapping.  On the right are 250,000 points generated with the
{\it nonequilibrium} map $B(q,p)$ where the comoving area changes by a factor of 2,
increasing about (1/3) of the time and decreasing about (2/3) of the time.  The
repellor ( shown at bottom left ) is a mirror image of the attractor ( $\pm p \rightarrow                                                  
\mp p$ ).  Whether one starts with the attractor or repellor the final distribution of
points invariably produces the attractor ( shown in the center and top right ).  The tops
and the bottoms of the diamond domains correspond to unstable fixed points.
}
\label{Figure1}
\end{figure}

See {\bf Figure 1} for the steady-state distribution correponding to the mapping $B(q,p)$.  Here
we consider a particular special case different to both the skinny and the classic maps.  In our case
the comoving density is doubled two-thirds of the time and halved one-third. The single-step dynamics
of the time-reversible nonequilibrium Baker Map $B$ converts a $(q,p)$ phase point to a new one,
$(q_{\rm new},p_{\rm new})$ always with a factor-of-two change in the comoving area.  The domain of the map
is diamond-shaped with a sidelength of 2, centered on the origin, $(q=0 \ ; \ p=0)$. The FORTRAN programming
necessary to each iteration of the map $B(q,p\rightarrow q_{\rm new},p_{\rm new})$ is as follows :
\begin{verbatim}                                                                                           
      if(q.lt.p-sqrt(2.0/9)) then ! [ expanding ]
         qnew = +(11.0/6)*q - (7.0/6)*p + sqrt(49.0/18)                                                    
         pnew = +(11.0/6)*p - (7.0/6)*q - sqrt(25.0/18)                                                    
      endif                                                                                                
                                                                                                           
      if(q.gt.p-sqrt(2.0/9)) then ! [ contracting ]                                                                          
         qnew = +(11.0/12)*q - (7.0/12)*p - sqrt(49.0/72)                                                  
         pnew = +(11.0/12)*p - (7.0/12)*q - sqrt( 1.0/72)                                                  
      endif                                                                                                
\end{verbatim}

In 1991, Bill Vance was interested in computing the relative probability of forward and reversed
trajectories using unstable periodic orbits\cite{b6}. He showed me that it is easy to confirm the
time-reversibility of the ``rotated'' Baker map $B(q,p \rightarrow q_{\rm new},p_{\rm new})$.
First, choose the point $(q,p)$ within the $2 \times 2$ diamond-shaped domain of {\bf Figure 1}
and compute $(q_{\rm new},p_{\rm new})$ using $B$. Then compute $B(q_{\rm new},-p_{\rm new})
\rightarrow (q,-p)$ and confirm that the resulting point $(q,-p)$ is indeed the time-reversed image
of the original $(q,p)$ point.

Another point of view from which to analyze the rotated map $B$ is ``statistical'', based on
distributions rather than on a single long dynamical trajectory. The statistical viewpoint describes
the evolution of the phase-space probability density due to transformations of the comoving
{\it area}, $dqdp \stackrel{B}{\rightarrow} dq_{\rm new}dp_{\rm new}$ . The map {\it doubles} the
comoving area whenever $q<p-\sqrt{(2/9)}$. $B$ {\it halves} the area when $q>p-\sqrt{(2/9)}$ :
$$
(11/6)^2 - (7/6)^2 = 2 \ ; \ (11/12)^2 - (7/12)^2 = (1/2) \ .
$$
The probability of observing $k$ expansions with $N-k$ contractions ( in any order ) in a total of
$N$ iterations of the map is the same as the biased random walk probability :
$$
(1/3)^k(2/3)^{N-k}{{N}\choose{k}} \ .
$$

\begin{figure}[h]
\centerline{\includegraphics[width=4.in,angle=-90,bb= 26 11 581 784]{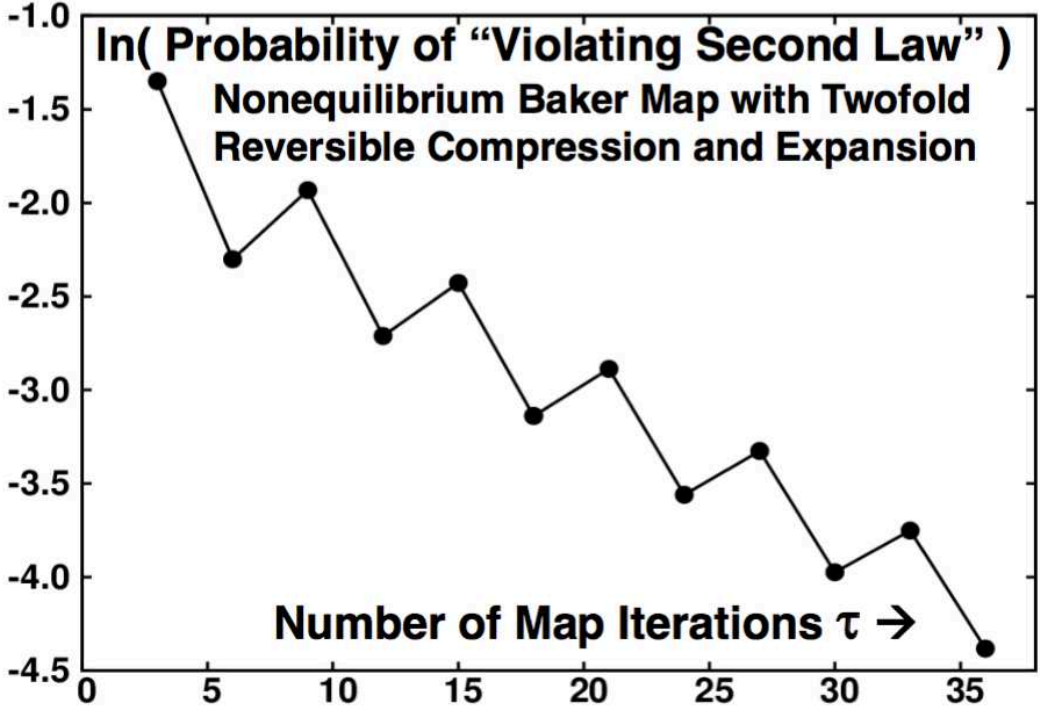}}
\caption
{
Iterations of the Baker map with twofold compression or expansion can yield negative
entropy production.  Here I show the probability of these ``Second Law Violations''
for from 3 to 36 iterations of the map, where all possible $2^\tau$ combinations of
compressions and expansions have been included and properly weighted.
}
\label{Figure2}
\end{figure}

For the most likely 12-step example the probability is $(1/3)^4(2/3)^8{{12}\choose{4}}$ while the
probability of the {\it reversed} walk, $(1/3)^8(2/3)^4{{12}\choose{4}}$ is sixteen times smaller.
For the $2^{36} = 68,719,476,736$ 36-step walks the probability of finding expansion rather than
contraction is still readily observable, and happens about one percent of the time.  In {\bf Figure 2}
we plot the probability of finding expansion, corresponding to negative entropy production, as a
function of the length of the observation window.  For windows evenly divisible by six the probability
varies roughly as $(2/3)^{(\tau/6)}$.

If we associate the logarithm of phase-space area with Gibbs' entropy the entropy changes by
$\pm k\ln (2)$ with each iteration of the map.  Because the region subject to halving is twice
as large as that subject to doubling, the {\it net} effect is an entropy {\it decrease} averaging
$-(k/3)\ln (2)$ per iteration.  Let us apply some steady-state thermodynamics.
We imagine a heat reservoir that extracts the entropy produced
by an iteration of the map and note that in the steady state that reservoir's entropy ( external to
the map ) increases at a nearly constant rate, $\simeq (k/3)\ln(2)$ per iteration of the map.  Although
this idea ( based on continuity ) might seem a bit suspect for a fractal distribution it appears to
be fully consistent with the ideas and results of Evans, Cohen, and Morriss.

In the Baker-Map case where the probability density is asymptotically uniform in the direction
parallel to the line $(q+p)=0$ and fractal in the perpendicular direction, parallel to the line
$(q-p)= 0$, the southwest third of the diamond necessarily comes to include (2/3)
of the natural measure.  The southwest third of the remainder ( see {\bf Figure 2} ) is a scale
model of the larger northeast third and contains (2/3) of the remaining measure, that is (2/9)
of the total. This scaling in the northeast direction continues on with each image of the original
smaller by a factor of (2/3) while containing (1/3) of the remaining measure. Thus the total area
of $2 \times 2 = 4$ is divided up as follows :
$$
(4/3) + (4/3)(2/3) + (4/3)(2/3)^2 + (4/3)(2/3)^3 + (4/3)(2/3)^4 + \dots = \frac{(4/3)}{1 - (2/3)}
\equiv 4 \ .
$$
The total ``natural measure'' is unity and corresponds to the following sum where the first term
corresponds to the southwest third of the domain :
$$
(2/3) + (2/3)(1/3) + (2/3)(1/3)^2 + (2/3)(1/3)^3 + (2/3)(1/3)^4 + \dots = \frac{(2/3)}{1 - (1/3)}
\equiv 1 \ .
$$ 
In our Baker Map (2/3) of the measure expands to the northwest, with a Lyapunov expansion of (3/2),
while being squeezed threefold in the perpendicular direction. Simultaneously (1/3) of the measure
expands to the southeast, moving mainly mostly east, with a Lyapunov expansion of 3. 
In the steady state the (2/3) of the measure halving in area perfectly balances the northeastern
motion of the remaining (1/3) which is doubling in area.

The fluctuation theorem for a window of $\tau$ steps states that the probability of converting work
to heat, divided by the ( illegal, according to the Second Law ) reversed process, is the exponential
of the entropy produced going forward in time :
$$
\ln \left[ \ {\rm prob}_{\rm forward}/{\rm prob}_{\rm backward} \ \right] = e^{+\sigma \tau} \ .
$$
For the Baker Map the probabilities ${\rm prob}_{\rm forward}(\tau)$ and ${\rm prob}_{\rm backward}(\tau)$
can be worked out analytically by noting that a sequence of steps forward in time is more probable than
its reverse by a factor of $e^{\sigma \tau}$.  There is an isomorphism linking the comoving expansions and
contractions of the Baker Map to a random walk in which steps to the right ( corresponding to compression )
are twice as likely as steps to the left ( corresponding to expansion ).  Let us look at the details of a
single example.

Consider the case $\tau = 6$ with $2^{6}$ outcomes for the entropy production ranging from
$(\Delta S/k) =-6\ln(2)$ with probability $(1/3)^{6}$ to $(\Delta S/k) = +6\ln(2)$ with probability
$(2/3)^{6}$.  Because this six-step mapping is equivalent to a random walk problem with left and right
probabilities of (1/3) and (2/3) the distribution for $(\Delta S/k)$ approximates a Gaussian with mean
value $\langle (\Delta S/k) \rangle = (\tau /3)\ln (2)$ and mean
squared value $[ \ (\tau/3)^2 +(8\tau/9) \ ][ \ \ln(2) \ ]^2$ :
$$
P_{-6} = \frac{1}{729} ; P_{-4} = \frac{6\cdot2}{729} ; P_{-2} = \frac{15\cdot4}{729} ;
P_0 = \frac{20\cdot8}{729} ; P_{+2} = \frac{15\cdot16}{729}; P_{+4} = \frac{6\cdot32}{729} ;
P_{+6} = \frac{64}{729} ,
$$ 
the binomial distribution for a biased random walk.  In the general case with a window $\tau$ the mean
entropy production is $\sigma = (\tau/3)$ and the mean of the squared entropy production is 
$$
\langle \sigma \rangle/(k\ln(2) = (\tau/3) \ ; \ \langle\sigma^2 \rangle /[ \ (k\ln(2) \ ]^2 =
(\tau/3)^2 + (8\tau/9) = \langle\sigma \rangle^2 + (8/3)\langle\sigma \rangle \ .
$$
From the Central Limit Theorem we see that the distribution approaches a Gaussian for large values of the 
entropy production 
$$
{\rm prob}(\sigma) \propto e^{(-3/16)(\sigma - \langle\sigma \rangle)^2/\langle\sigma \rangle}
$$
The biased random walk problem, which satisfies the Fluctuation Theorem precisely, is a useful model for
introducing students to the voluminous literature on this subject.\cite{b7}

\pagebreak

\end{document}